\documentclass[]{aa}
\usepackage{graphicx}
\usepackage{longtable}

\newcommand{\grp}    {${\rlap.}^{\circ}$}
\newcommand{\pri}    {${\rlap.}^{\prime \prime}$}
\newcommand{\rl}     {${\rlap.}^{s}$}

\newcommand{\ltsima} {$\; \buildrel < \over \sim \;$}
\newcommand{\simlt}  {\lower.5ex\hbox{\ltsima}}            
\newcommand{\gtsima} {$\; \buildrel > \over \sim \;$}
\newcommand{\simgt}  {\lower.5ex\hbox{\gtsima}}            

\begin{document}

\title{Deep radio images of the HEGRA and Whipple TeV sources in the Cygnus OB2 region}

\author{Josep Mart\'{\i}\inst{1}, Josep M. Paredes\inst{2}, Ishwara Chandra C. H.\inst{3},
Valent\'{\i} Bosch-Ramon\inst{4}}

\offprints{J. Mart\'{\i}}

\institute{
Departamento de F\'{\i}sica, Escuela Polit\'ecnica Superior, Universidad de Ja\'en,
Las Lagunillas s/n, 23071 Ja\'en (Spain)
\email{jmarti@ujaen.es}
\and
Departament d'Astronomia i Meteorologia, Facultat de F\'{\i}sica, Universitat
de Barcelona, Mart\'{\i} i Franqu\`es, 1, 08028 Barcelona (Spain)
\email{jmparedes@ub.edu}
\and
National Center for Radio Astrophysics, TIFR, P. B. No. 3, Ganeshkhind, Pune - 7 (India)
\email{ishwar@ncra.tifr.res.in}
\and
Max Planck Institut f\"ur Kernphysik, Saupfercheckweg 1, Heidelberg 69117 (Germany)
\email{vbosch@mpi-hd.mpg.de}
}

\date{Received 25 April 2007/ Accepted 21 May 2007}

\authorrunning{Mart\'{\i} et al.}

\abstract
{The modern generation of Cherenkov telescopes has revealed a new population of gamma-ray sources in the Galaxy.
Some of them have been identified with previously known X-ray binary systems while other remain without clear
counterparts a lower energies. Our initial goal here was reporting on extensive radio observations 
of the first extended and yet unidentified source, namely TeV~J2032+4130. This object was originally 
detected by the HEGRA telescope in the direction of the Cygnus OB2 region and its nature has 
been a matter of debate during the latest years. The situation has become more complex
with the Whipple and MILAGRO telescopes  new TeV detections in the same field which 
could be consistent with the historic HEGRA source, although a different origin cannot be ruled out.}
{We aim to pursue our radio exploration of the TeV~J2032+4130 position that we initiated in a previous paper but
taking now into account the latest results from new Whipple and MILAGRO TeV telescopes.
The data presented here are an extended follow up of our previous work.}
{Our investigation is mostly based on interferometric radio observations with 
the Giant Metre Wave Radio Telescope (GMRT) close to Pune (India) and
the Very Large Array (VLA) in New Mexico (USA).
We also conducted near infrared observations with the 3.5 m telescope and the OMEGA2000 camera at the 
Centro Astron\'omico Hispano Alem\'an (CAHA) in Almer\'{\i}a (Spain).}
{We present deep radio maps centered on the TeV~J2032+4130 position at different wavelengths. In particular, our 49 and 20 cm 
maps cover a field of view larger than half a degree that fully includes the Whipple position and the peak of MILAGRO emission.
Our most important result here is a catalogue of 153 radio sources detected at 49 cm within the GMRT antennae primary beam
with a full width half maximum (FWHM) of 43 arc-minute. Among them, peculiar sources inside the Whipple error ellipse
are discussed in detail, including a likely double-double radio galaxy and a one-sided jet source of possible blazar nature.
This last object adds another alternative counterpart possibility to be considered
for both the HEGRA, Whipple and MILAGRO emission.
Moreover, our multi-configuration VLA images reveal the non-thermal extended emission previously reported by us
with improved angular resolution. Its non-thermal spectral index is also confirmed thanks
to matching beam observations at the 20 and 6 cm wavelengths.
}
{}
\keywords{Radio continuum: stars -- X-rays: binaries -- Galaxies: active -- Gamma rays: observations}

\maketitle

\section{Introduction}

The unidentified TeV gamma-ray source 
TeV~J2032+4130 in the Cygnus OB2 region was originally reported by the
HEGRA array of imaging Cherenkov telescopes (\cite{aha2002}). Together with
HESS J1303$-$631  (\cite{aha2005a}), it is one of the two first extended TeV sources
without a clear counterpart at lower energies. In an attempt to explain the nature
of these unidentified extedend gamma-ray sources,
several TeV emission models based on hadronic processes instead of leptonic ones have
flourished during the past years (e.g. \cite{t2004, br2005}). However, 
the only consensus so far achieved is that these
mysterious objects are likely to be galactic sources based on their Galactic Plane location
and extended
nature of their TeV emission.

During the latest years, several authors have undertaken deep observational efforts to try to identify
potential counterparts at other wavelengths (\cite{butt2003, butt2006, mu2003}).
No clear success has been achieved so far at both X-ray, infrared and specially at radio wavelengths
where interstellar extinction should be minimal. At present, the deepest radio exploration has been reported
by \cite{par2007} based on new observations with the Giant Meter Wave Radio Telescope (GMRT)
in Pune (India) and new and archive observations with the Very Large Array (VLA) in New Mexico (USA). As a result of this work,
an extended diffuse radio emision (likely of non-thermal nature) and a remarkable population of compact radio sources
were detected in coincidence with the Center of Gravity (CoG) and few arc-minute radius of the HEGRA
extended emission. X-ray and near infrared counterparts have been identified for some of these compact radio sources.
Their nature ranges from stellar to likely extragalactic sources but none turned out to be an uncommon or
peculiar object from which TeV emission could be suspected. 

More recently, new observations conducted with the Whipple Cherenkov telescope
(\cite{kono2007}) have resulted in the detection of an extended TeV source
displaced several arc-minutes NE from the HEGRA position 
but yet consistent with it when all sources of error are considered. The contents of the Whipple error box,
corresponding to the (significant) excess ellipse,
has been studied by \cite{butt2007} 
who report on non-thermal extended radio emission consistent with that found by \cite{par2007}.
A possible supernova remnant (SNR) interpretation is tentatively proposed for it. They also discuss a conspicuous
double-lobed jet-source, now within the new Whipple ellipse and previously noted by \cite{butt2006} and \cite{par2007}.
But the latest complication to this puzzle has been added by the water Cherenkov telescope MILAGRO as
reported by \cite{abdo2006}.
The map of the region obtained by the MILAGRO collaboration indicates the presence of a conspicuous
TeV emitter whose position is consistent with both the HEGRA and Whipple sources. 
However, the association of any of the Whipple and MILAGRO sources with TeV~J2032+4130, although possible,
is at present unclear
and the existence of different TeV sources in the field cannot be ruled out.
 
In this context, we present additional radio data covering both the HEGRA, Whipple and MILAGRO peak positions. 
Firstly, the GMRT observations in \cite{par2007} (covering the HEGRA position only)
are now reported for the whole field of view of the GMRT antennae primary beam.
Secondly, this paper is also based on new VLA observations that we combine with the previous VLA archive data for high fidely
mapping of extended emission.
Our final maps provide the highest sensitivity and angular resolution radio images so far available 
for the TeV sources in the Cygnus OB2 region. In particular, the full analysis of GMRT data has resulted
in a catalogue of 153 radio sources reliably detected in the field which we consider to be the most important contribution of this work. 
The availability of such information in the radio will become very useful in a near future when improved Cherenkov 
and GLAST observations will be able to further decrease
the uncertainty of the TeV position. The true radio counterpart is then likely to emerge among the many radio
sources reported in this paper.

\section{GMRT radio observations and results}

\begin{figure*}
   \centering
\vspace{18cm}
\includegraphics{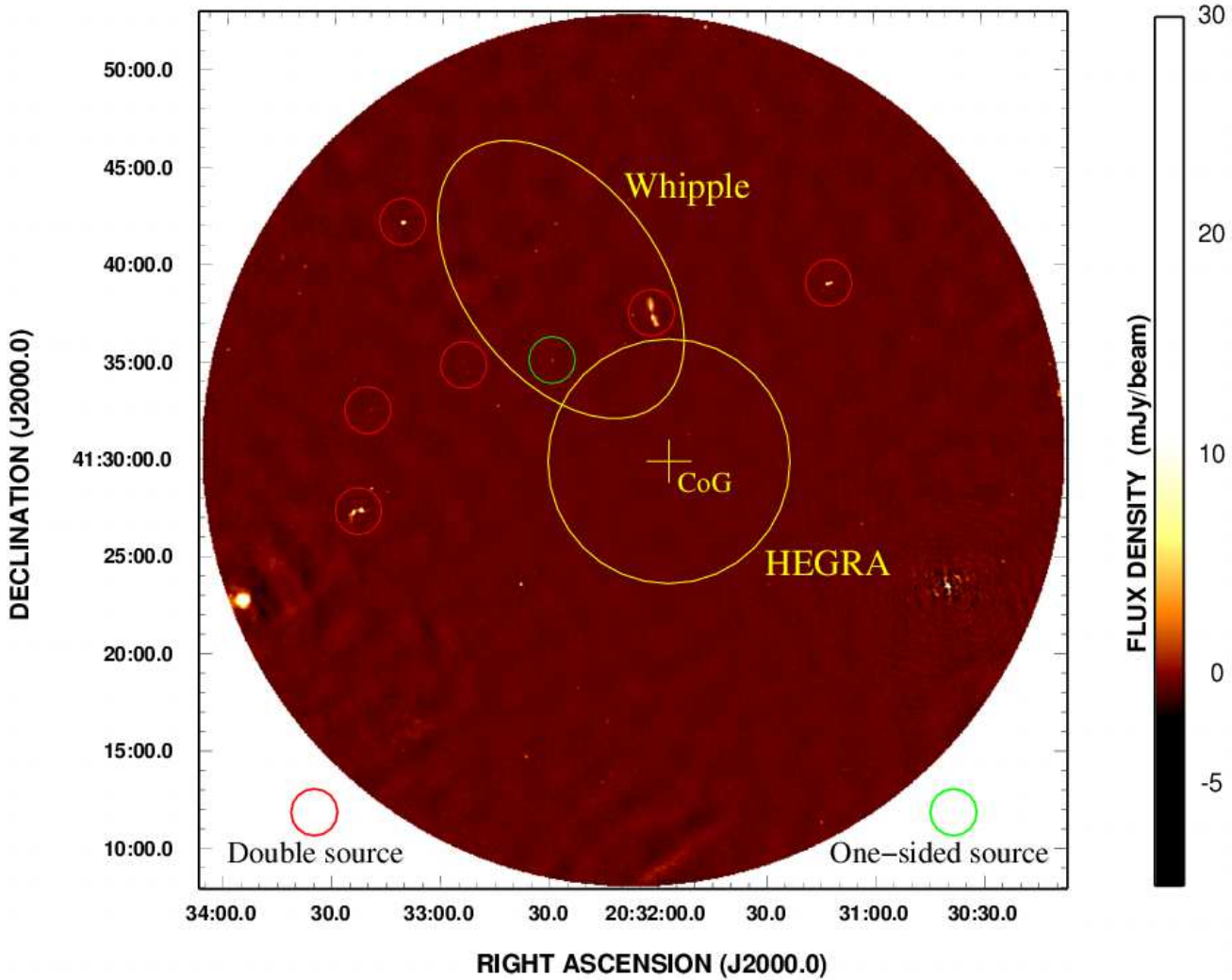}
      \caption{
Wide field radio map of the HEGRA and Whipple TeV sources and its surroundings in the Cygnus OB2 region
based on multi-epoch GMRT observations at 49 cm
corrected for primary beam response up to 50\%.
The CoG of HEGRA TeV emission and its statistical error are
indicated by the central cross, while the central circle ilustrates
the $1\sigma$ radius of its extended angular size.
The elongated ellipse approximately represents the Whipple location from \cite{kono2007},
believed to correspond to the same TeV source and
clearly offset towards the North East.
The color scale is logarithmic and shown on the right side. The synthesized beam amounts to
5\pri 00 $\times$ 4\pri 75, with position angle of 33\grp 0.
It is not shown because it would appear just as a small dot at this scale.
Resolved double and one-sided jet sources are indicated by color circles red and green, respectively.
The rms noise in the central region is about 90 $\mu$Jy beam$^{-1}$.
}
      \label{gmrt_map}
   \end{figure*}

The GMRT observations were carried out at 610 MHz (49 cm wavelength) on 2005 July 9 and September 1
according to the log in Table \ref{gmrtlog}. They were preliminary reported by \cite{par2007}, but
limited to the HEGRA circle contents. We refer the reader to this reference for details on the GMRT 
observing, calibration and data processing techniques. Here, we will provide a full account of our GMRT observing runs 
including all radio sources within the FWHM of the primary beam of individual antennae.
The corresponding wide-field map (about 43 arc-minute wide) is presented in Fig. \ref{gmrt_map}. Such field of view  
was conservatively adopted for two reasons mainly. Firstly, it safely accomodates the location uncertainty of both
the HEGRA and Whipple sources. Secondly, it avoids known problems in the modelling of the primary
beam response beyond the 50\% level (see Appendix 2 in \cite{garn2007}).

Many radio sources are detected in Fig. \ref{gmrt_map} when closely inspecting the fits file.
We used the automated extracting procedure SAD of the AIPS package in order to produce
a list of objects with peak flux density higher than about five times the local noise
after primary beam correction. The SAD output was inspected visually and all source detections
believed not to be real (i.e. deconvolution artifacts near bright sources) were deleted by hand.
The resulting list, considered to be very reliable although not complete at lowest flux density levels, 
contains 153 radio sources. We present it as a catalogue in
Table \ref{list} of the online material accompanying this paper. A sample of this table is included
here for illustration purposes of our GMRT catalogue. First and second columns provide the J2000.0 position in right ascension order.
The third and fourth columns give the peak ($S_{\nu}^{\rm Peak}$) and integrated ($S_{\nu}^{\rm Integ}$) flux densities. 
Finally the fifth, sixth and seventh
columns contain the result of an elliptical Gaussian fit as returned by the same SAD task. This is an apparent
angular size, i.e., not a deconvolved one. All errors are given in parentheses.

Very extended sources with $S_{\nu}^{\rm Integ}/S_{\nu}^{\rm Peak}$\gtsima$6$, such as some double-lobed radiogalaxies,
are not included in Table \ref{list} and their detailed study is reserved for
future papers. However, we do pay attention below to some of these
objects that we consider relevant to mention for our goals of TeV source identification.

\begin{table}
\caption[]{\label{vlaobs} Log of GMRT observations used in this paper \label{gmrtlog}.}
\begin{tabular}{cccccc}
\hline
\hline
Date \&     &  $\lambda$ & Anten.          & IF    & IF     & Visibility \\
Project     &    (cm)    & used            & \#    & Width  &     \#     \\
 Id.        &            &                 &       & (MHz)  &            \\
\hline
2005 Jul 9  & 49    &  all        &   1    & 13 (LSB)    &  340605 (LSB) \\
(08JMP01)   &       &             &        & 16 (USB)    &  340170 (USB) \\
            &       &             &        &             &               \\
2005 Sep 1  & 49    &  all        &   1    & 13 (LSB)    &  276939 (LSB) \\
(08JMP01)   &       &             &        & 16 (USB)    &  276939 (USB) \\
\hline
\hline
\end{tabular}
\end{table}


\begin{table*}
{\bf Table 3 (sample).} Catalogue of GMRT sources detected at 610 MHz (full catalogue in the online section). \\
\begin{tabular}{ccccccc}
\hline
\hline
$\alpha_{\rm J2000.0}$ & $\delta_{\rm J2000.0}$  &  $S_{\nu}^{\rm peak}$ & $S_{\nu}^{\rm integ}$  &  $a$    & $b$      &   P.A.    \\   (hms)               &          (dms)          & (mJy beam$^{-1}$)     & (mJy)               &  (arcsec)  & (arcsec) & (degree)  \\\hline
20 30 08.779(0.004) &   +41 33 16.60(0.03) &   8.18(0.11) &  17.30(0.32) &  8.47(0.11) &  5.93(0.08) &  83.0(001) \\
20 30 09.090(0.015) &   +41 33 26.63(0.21) &   2.31(0.10) &   9.23(0.51) & 11.16(0.51) &  8.51(0.39) &   4.0(007) \\
20 30 11.448(0.015) &   +41 34 52.19(0.15) &   2.14(0.11) &   4.95(0.34) &  8.29(0.42) &  6.63(0.33) &  68.5(009) \\
20 30 13.065(0.066) &   +41 35 29.88(0.59) &   0.52(0.11) &   0.94(0.29) &  9.58(2.02) &  4.51(0.95) &  54.5(010) \\
20 30 14.295(0.040) &   +41 24 54.89(0.52) &   0.66(0.11) &   1.30(0.30) &  7.48(1.24) &  6.26(1.03) &  18.5(037) \\
\hline
\end{tabular}
\end{table*}

\section{VLA radio observations and results}

   \begin{figure*}
   \centering
\vspace{16.75cm}
\includegraphics{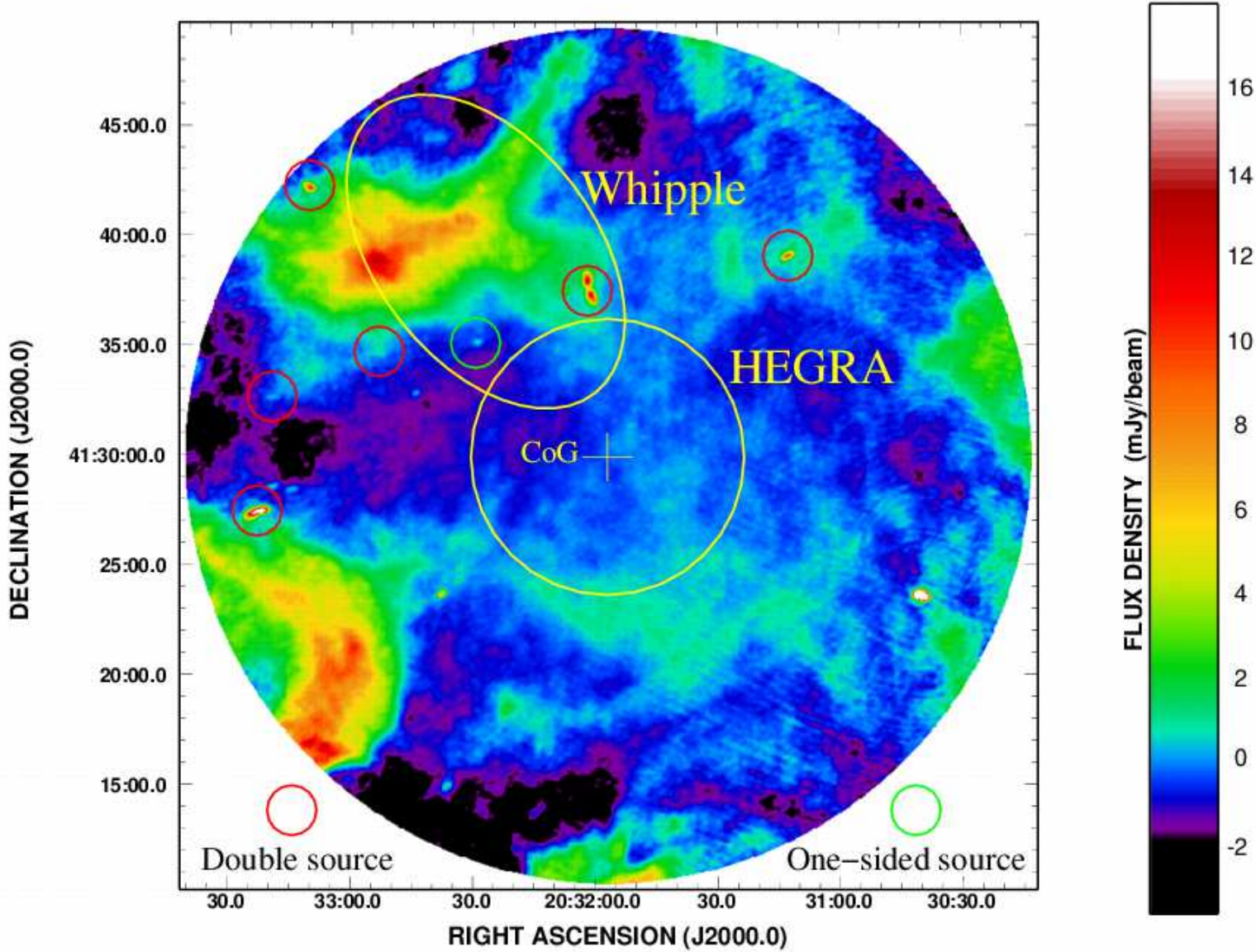}
      \caption{
VLA 20 cm map of the HEGRA location for the unidentified TeV~J2032+4130 source.
The location of the HEGRA and Whipple sources are superimposed on it
with the same symbols of Fig. \ref{gmrt_map}.
The 20 cm radio data shown here is a
natural weight map made using VLA data taken in C and D configuration and combined.
The circular map boundary is due to the primary beam correction up to a 30\% response, thus
increasing the rms noise toward the edges.
The vertical wedge indicates the brightness level in mJy beam$^{-1}$ within limits
adapted to better enhance the most interesting features in the field.
The synthesized beam corresponds to an ellipse of
16\pri 97$\times$13\pri 33, with position angle of $-$63\grp 85. Again, it is not shown because
it would appear as a very small dot at this scale. Color circles indicate sources
with resolved morphology mainly based on GMRT observations discussed later in the text.}
      \label{vla_c+d_20cm}
   \end{figure*}

\begin{table}
\caption[]{\label{vlalog} Log of VLA observations used in this paper.}
\begin{tabular}{cccccc}
\hline
\hline
Date  \&  &     $\lambda$   & VLA    & IF    & IF     & Visibility \\
Project   &       (cm)      & configuration  & \#    & Width  &     \#     \\
  Id.     &                 &        &       & (MHz)  &            \\
\hline
2003 Apr 29 & 20 &  D & 2 & 50 & 392125       \\
 (AB1075)   &  6 &  D & 2 & 50 & 339270       \\
            &    &    &   &    &              \\
2006 Nov 21 & 20 &  C & 2 & 50 &  70317       \\
  (AM871)   &  6 &  C & 2 & 50 &  93624       \\
            &    &    &   &    &              \\
2006 Nov 28 & 20 &  C & 2 & 50 &  76069       \\
  (AM871)   &  6 &  C & 2 & 50 &  92433       \\
\hline
\hline
\end{tabular}
\end{table}

The VLA observations were carried out during November 2006 at the 20 and 6 cm
wavelengths with the array being in its C configuration.
The corresponding log is given in Table \ref{vlalog}. For completeness
of the VLA data used in this work, Table \ref{vlalog} also contains
other archive observation obtained by other authors in the more compact D configuration of the array and used by us.
All VLA data were processed following standard procedures within the AIPS package of NRAO.
3C48 was observed as amplitude calibrator at all wavelengths.
Phase calibration was achieved by frequent observations of
2052+365 and 2007+404 at 20 and 6 cm, respectively. Phase self-calibration worked
satisfactorily at 20 cm while at 6 cm the brightest sources were not strong enough to enable it.

   \begin{figure*}
   \centering
\vspace{16.5cm}
\includegraphics{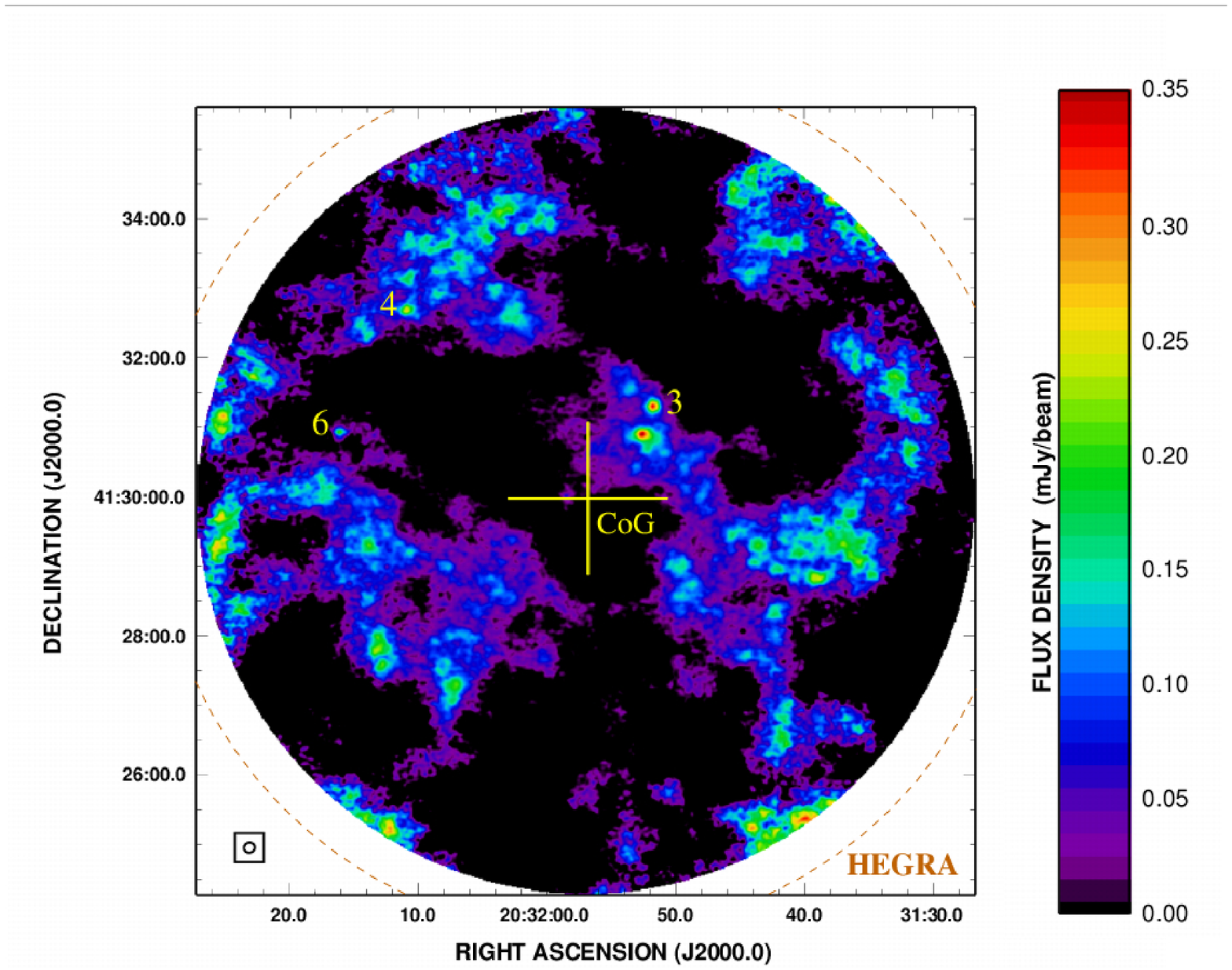}
      \caption{
VLA 6 cm map of the HEGRA location for the unidentified TeV~J2032+4130 source.
This is a natural weight map using the C and D configuration of the array
and combined in the $uv$ plane.
The HEGRA $1\sigma$ radius and CoG are the plotted as in previous figure.
The vertical wedge indicates the brightness level in mJy beam$^{-1}$.
The synthesized beam is indicated by the bottom left ellipse and corresponds to
9\pri 55$\times$8\pri 69, with position angle of $-$74\grp 44.
Primary beam correction has been also applied limited to a
30\% response. This provides a field of view almost as large as the
HEGRA circle. The Id. numbers correspond to GMRT radio sources reported by
\cite{par2007} which are also detected here at 6 cm.
}
      \label{vla_c+d_6cm}
   \end{figure*}

   \begin{figure*}
   \centering
\resizebox{\hsize}{!}{\includegraphics[angle=0]{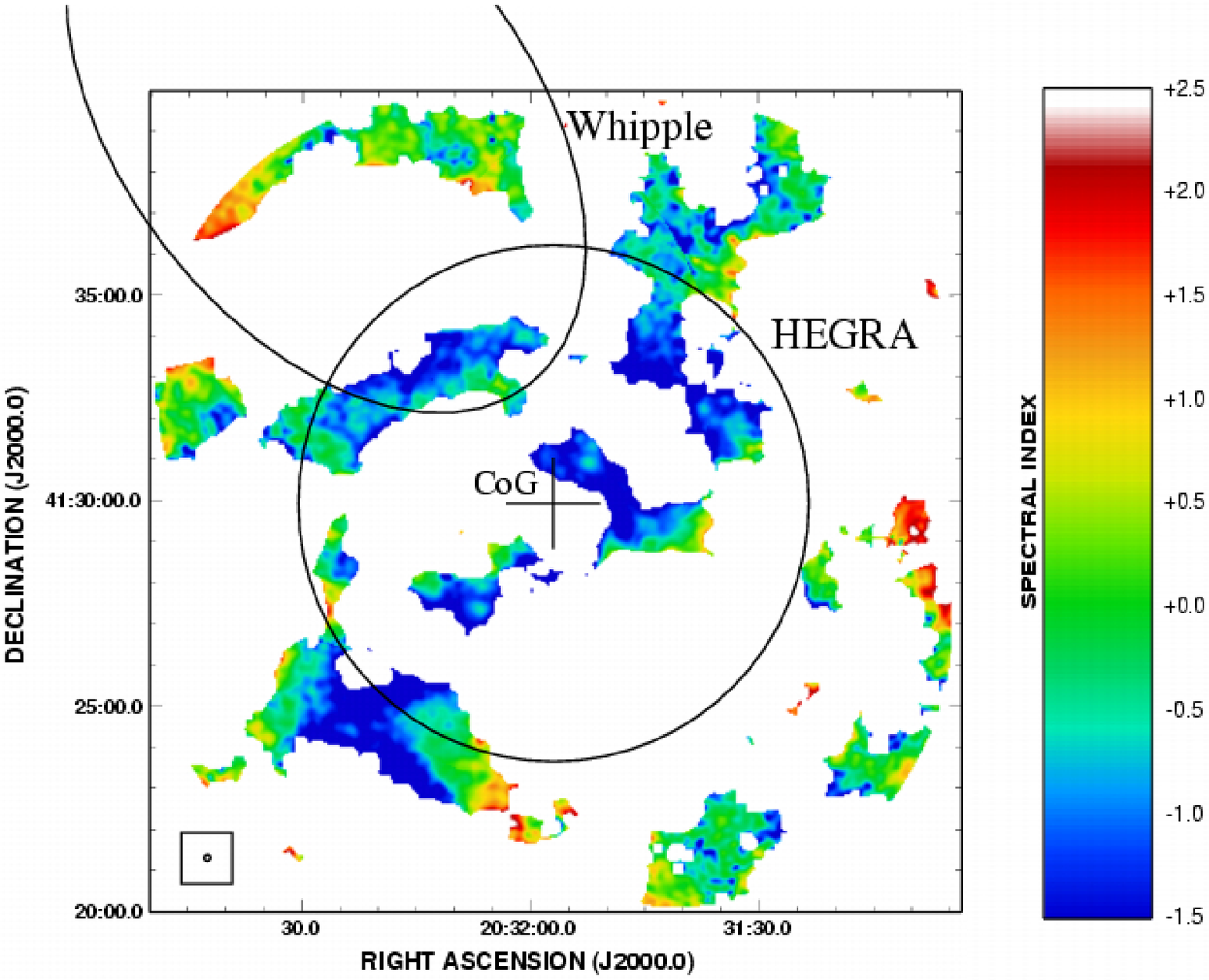}}
      \caption{
Spectral index map computed by combining matching beams maps
made from  VLA 6 cm data in D configuration and VLA 20 cm data in C configuration.
The restoring beam used was a circular Gaussian with $15^{\prime\prime}$ diameter
at both frequencies shown at the bottom left corner.}
      \label{spix}
   \end{figure*}

The purpose of our VLA observations was twofold. Firstly, the VLA C configuration run was
conducted with the idea of combining it (in the Fourier plane)
with the other D configuration runs available from the archive.
A map computed using C+D data is expected to provide both improved angular resolution and sensitivity to
extended radio emission in order to better image possible radio counterparts to the TeV source.
The result of combining C and D configuration data, using the aips task DBCON, is presented in
Figs. \ref{vla_c+d_20cm} and \ref{vla_c+d_6cm}.
These multi-configuration images display a significant
amount of detail both in compact and diffuse emission. Primary beam correction has been applied to them
limited to a 30\% response of the primary beam for higher fidelity purposes.

Secondly, a C configuration run can also be used to provide true matching beam maps
at both 20 and 6 cm. Indeed, observing at 20 cm in C configuration provides
almost the same synthesized beam as the 6 cm archive data in D configuration.
When these data are appropriately combined, spectral indices can be determined in a more
reliable way and specially for extended radio emission.
Figure \ref{spix} displays the result of combining the matching beam maps quoted above (not shown here)
in spectral index mode using the aips task COMB.
The field of view covered is larger than a single 6 cm primary beam because the archive VLA D configuration data
consisted of mosaiced pointings at this wavelength.

\section{CAHA near infrared observations and results}

On 2005 April 29, we observed the field of TeV~J2032+4130 in the near infrared $Ks$-band (2.2 $\mu$m) using the
Calar Alto 3.5~m telescope and the OMEGA2000 camera at CAHA. Only five frames could be obtained close
in time to the astronomical twilight. Therefore, the sky subtraction is not as perfect as desirable and this limits somewhat
the quality of the data. Nevertheless, they are comparable or even better than the 2MASS with a
limiting magnitude of $K_s=15.5$.  The relevant $Ks$-band fields are displayed as a
background gray scale in Figs. \ref{rgal} and \ref{gmrt_map_zoom}.
These CAHA observations have been partially published already in \cite{par2007} to whom
we refer the reader for futher details.

\section{Discussion}

\subsection{The GMRT catalogue of the region}

Possibly the most important contribution of this paper is the GMRT source catalogue 
compiled in Table \ref{list}. Having this information in hand will likely become a useful
reference for radio counterpart identification when
future Cherenkov and GLAST observations will be able to narrow
the location of the HEGRA, Whipple and MILAGRO sources. 
The distribution of radio sources with flux density is given by the histogram
in Fig. \ref{histo}. We note here that the number of sources would become significantly higher
by $\sim 10^2$ if objects at the 3-4 $\sigma$ level were accepted in the search. 

\begin{figure}
   \centering
\resizebox{\hsize}{!}{\includegraphics[angle=0]{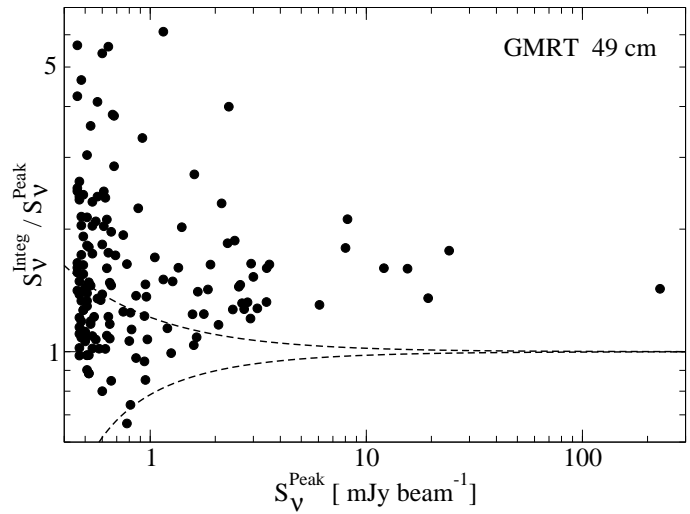}}
      \caption{Ratio of the integrated flux density $S_{\nu}^{\rm Integ}$ over the
peak flux density $S_{\nu}^{\rm Peak}$ for all GMRT radio sources in Table \ref{list} catalogue.
Dashed lines are the envelope of loci where sources are statistically consistent
with being unresolved (from \cite{bondi2007}).}
      \label{ratio}
   \end{figure}

In Fig. \ref{ratio} we have plotted the ratio $S_{\nu}^{\rm Integ}/S_{\nu}^{\rm Peak}$
as a function of peak flux density. The additional dashed lines correspond to the envelope equation
\begin{equation}
S_{\nu}^{\rm Integ}/S_{\nu}^{\rm Peak} = 0.85^{-(1.2/S_{\nu}^{\rm Peak})}  \label{envel}
\end{equation}
and its mirror image with respect to the unity value in the vertical axis. This envelope has been 
adopted from \cite{bondi2007} based on their deep source counts using the GMRT at 610 MHz. Sources within 
the two envelope branches are statistically consistent with being compact, or unresolved, while those above
can be classified as resolved objects. Out of the 153 objects in Table \ref{list}, about 40\% are compact
and 60\% appear resolved (61 and 92 radio sources, respectively). The dominance of resolved radio sources
is clearly connected the excellent GMRT angular resolution provided by its longest baselines.

The richness of catalogues based on a wide-field instrument such as GMRT allows the prospective of interesting
cross-identifications of its 610 MHz radio sources with other wavelength domains. Of course,
this task is not limited only to the TeV sources that originally motivated our work. As an example 
of quick data mining using the Table \ref{list} catalogue, the object marked with footnote $(^d)$
is very likely the lowest frequency detection of the contact binary system Cyg OB2 No. 5 (see e.g. \cite{con1997}
for a detailed optical-radio study). Additional hot massive stars worth to be investigated 
could also be present in Table \ref{list}. In particular, sources marked with footnotes $(^e)$ and $(^f)$
are very close in position to stars Cyg OB2 No. 12 and 9, respectively.

\subsection{Peculiar radio sources in the HEGRA/Whipple field}

The Whipple ellipse contains some interesting objects outside its overlap with the HEGRA circle but not too far
from its edge radius. Their existence has already been noted in the past based on survey data.
One of them is the very extended ($\sim10^{\prime}$) radio source 080.45+01.07 (\cite{furst90}) previosuly
suspected as a possible large scale radio lobe of the nearby microquasar Cygnus X-3 (\cite{marti2000}),
but now believed to be a mere HII region of thermal emission nature. The other is the prominent
double-lobed source mentioned by \cite{butt2006} and \cite{par2007}, also known as NVSS J203201+413722.
None is included in our catalogue due to their very extended nature.

Both objects are imaged here with improved detail. However, the most detailed view of these and other sources in the field
comes from the GMRT 49 cm map in Fig. \ref{gmrt_map} with about $5^{\prime\prime}$ angular resolution.
Thanks to this resolving power, the number of double sources existing in the HEGRA and Whipple field increases significantly
to at least six objects (see again Fig. \ref{gmrt_map}). Such abundance reduces the peculiarity of the 
double-lobed source as a counterpart candidate to be responsible for the TeV~J2032+4130/Whipple emission. 
Moreover, the GMRT map also reveals a third remarkable object, i.e., an unknown
one-sided jet source also consistent with the Whipple ellipse and 
not far from the HEGRA circle. We discuss more in detail on these radiosources in 
the subsections below.

\subsection{NVSS J203201+413722: a likely radio galaxy with an optically-thick compact core}

\begin{figure*}
   \centering
\resizebox{\hsize}{!}{\includegraphics{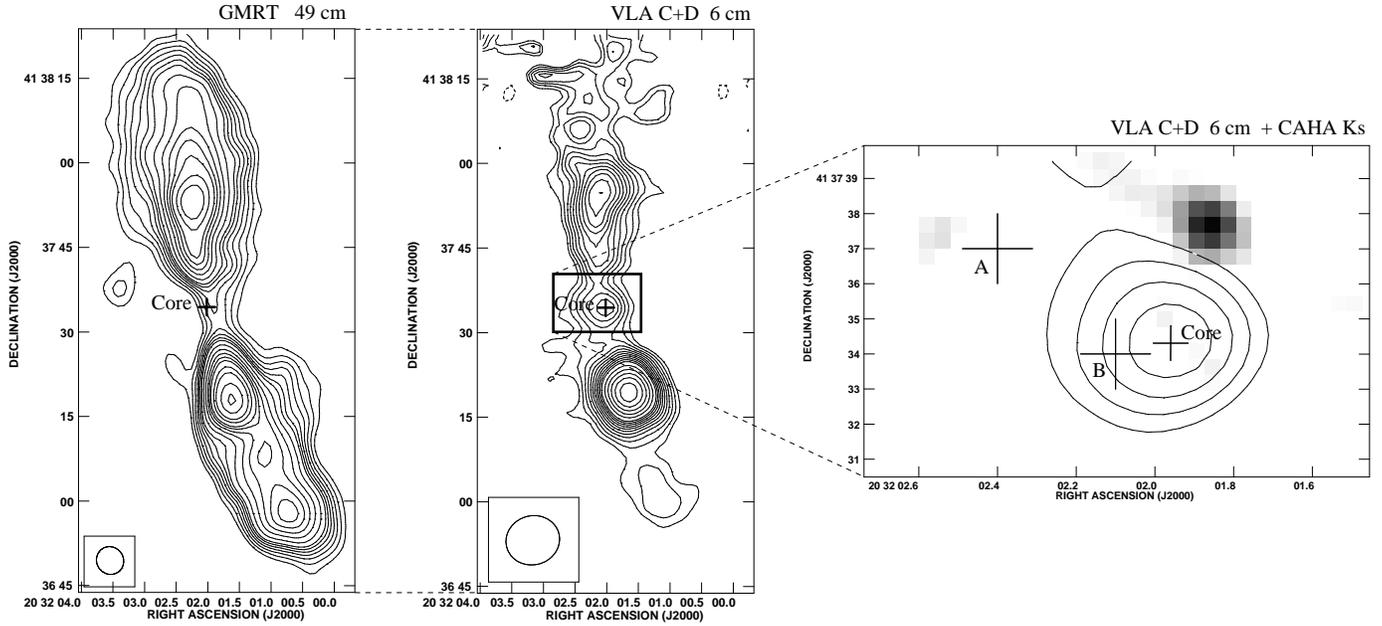}}
      \caption{
{\bf Left.} GMRT radio map of the proposed radiogalaxy at the 49 cm wavelength computed using
a ROBUST value of 0 and excluding baselines shorter than 1.1 k$\lambda$. Primary beam correction
has been also applied. Contours correspond to $-4$, 4, 6, 8, 10, 12, 14, 16, 20, 25, 30, 35, 40, 50, 60,
80 and 100 times 0.12 mJy beam$^{-1}$, the rms noise. The synthesized beam is shown as an ellipse 
at the bottom left corner and is equivalent to 5\pri 00 $\times$ 4\pri 75, with position angle of
33\grp 0.
{\bf Center.} The same field as in previous panel but observed with the VLA at 6 cm. This is a pure
natural weight, primary beam corrected VLA map 
produced by combining $uv$ data taken in the D and C configurations of the array.
A central thick cross marks the position of a compact component proposed to be the radiogalaxy core.
This component is not visible in the left GMRT map also with the same cross symbol for easier comparison.
Contours are $-4$, 4, 5, 6, 7, 8, 9, 10, 11, 12, 13, 14, 16, 18, 20, 22 and 24 times 0.14 mJy beam$^{-1}$.
The synthesized beam is similarly shown and corresponds to 
9\pri 55$\times$8\pri 69, with position angle of $-$74\grp 44.
{\bf Right.} Zoomed map of the proposed radio core made using the same VLA 
data of the central panel but using a ROBUST factor
of zero for enhanced angular resolution. Contours are $-4$, 4, 5 and 6 times
0.17 mJy beam$^{-1}$, the rms noise, with the synthesized beam 
being 5\pri 82$\times$4\pri 71 with position angle of $-$82\grp 65 (not shown for clarity reasons).
Both the core location and the Chandra X-ray sources 
from \cite{butt2007} are accurately marked here
as thin crosses whose size reflects their positional
uncertainty. The background gray scale indicates near infrared $K_s$-band emission as observed
with the Calar Alto 3.5~m telescope and the OMEGA2000 camera.}
      \label{rgal}
   \end{figure*}

A detailed GMRT view of this double-lobed source inside the Whipple ellipse is presented in the left panel of Fig.
\ref{rgal}. Its morphology is clearly reminiscent of a double-double radio galaxy which has experienced different
epochs of jet activity (see e.g. \cite{sch2000}).
The same field is also shown at 6 cm from our multi-configuration VLA observations.
The primary beam correction for this object at 6 cm is very high since it is located
below the 30\% level response. However, we are confident that its overall 6 cm morphology is not strongly affected
because it is a relatively strong source.

The most interesting issue about NVSS J203201+413722 is revealed by the detection of 
a compact component in the 6 cm map located at a J2000.0 position 
$20^h32^m$01\rl 96 and $+41^{\circ}37^{\prime}$34\pri 3 (error $\pm$0\pri 5 in each coordinate)
and with a flux density of $1.1\pm0.2$ mJy.
The fact that we do not see it at 20 nor 49 cm, above four times the local rms noise,
implies an optically-thick spectral index of about $\alpha \geq +0.4$. This is 
possibly due to synchrotron self absorption. \cite{butt2007} have reported marginal Chandra X-ray detections 
in the vicinity of NVSS J203201+413722 and their source B is consistent with this optically-thick component.
All these facts, together with the central symmetric location of the component with respect to the lobes,
strongly suggests that we have identified the true location of the central engine powering the 
rest of the non-thermal jet source. It is at this precise location where the search for a near infrared counterpart should
be conducted in order to test its galactic or, most likely, extragalactic radio galaxy nature.
The zoom in the right panel of Fig. \ref{rgal} also contains in gray scale our CAHA OMEGA2000  observations that
reveal no counterpart to the core component brighter than the OMEGA2000 limiting magnitude.

   \begin{figure}
   \centering
\resizebox{\hsize}{!}{\includegraphics[angle=0]{histo.eps}}
      \caption{Histogram showing the
distribution of peak flux densities for GMRT radio sources in Table \ref{list} catalogue.}
      \label{histo}
   \end{figure}

   \begin{figure*}
   \centering
\vspace{8.2cm}
\includegraphics{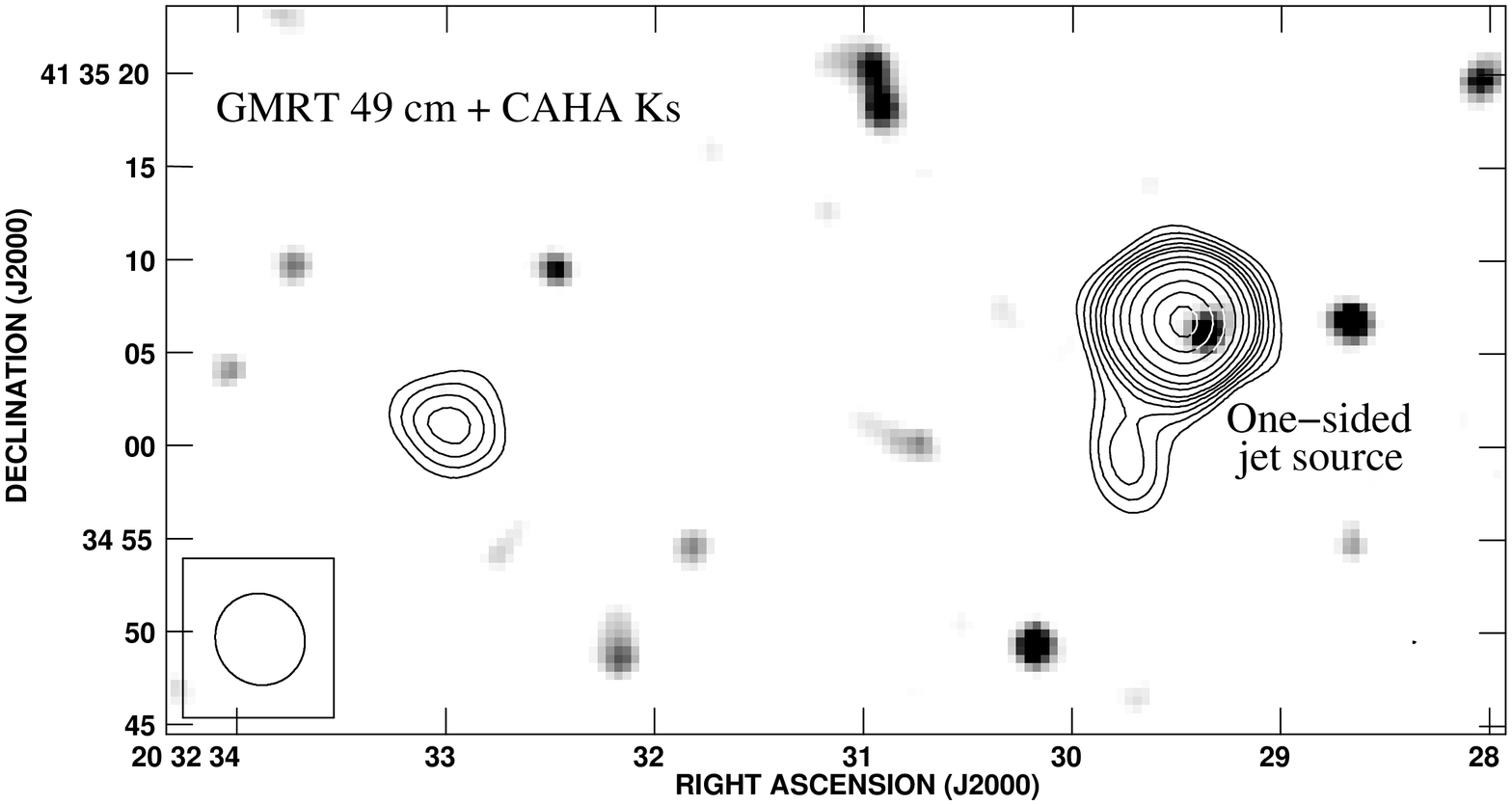}
      \caption{
Contour map of the one-sided jet source detected by the GMRT inside the Whipple ellipse
for TeV~J2032+4130. Contour levels shown correspond to $-3$, 3, 4, 5, 6, 7, 8, 10, 12, 15, 20, 25 and 30 times
the rms noise of 90 $\mu$Jy beam$^{-1}$.
This plot is a zoomed region of the same GMRT image as in Fig. \ref{gmrt_map} and with the same synthesized beam
(shown at the bottom left corner).
The background gray scale shows the $K_s$-band emission as observed
with the Calar Alto 3.5~m telescope and the OMEGA2000 camera. An obvious stellar-like counterpart candidate
is visible in the near infrared for this peculiar radio source.
}
      \label{gmrt_map_zoom}
   \end{figure*}

\subsection{A blazar candidate counterpart for the Whipple emission?}

A source with a clearly asymmetric one-sided jet is located inside the Whipple ellipse 
at J2000.0 coordinates $20^h32^m$29\rl 48 and $+41^{\circ}35^{\prime}$06\pri 7,
with a peak flux density of 2.8 mJy (see Table \ref{list} for details).  
This object, included in the GMRT catalogue, is labelled with a green circle in Fig. \ref{gmrt_map}
and a more detailed contour plot is presented in Fig. \ref{gmrt_map_zoom}.
Jet one-sideness is usually interpreted in terms of strong Doppler boosting of a relativistic
jet flowing close to the line of sight. This is the main fingerprint 
of blazar active galactic nuclei (AGN) which are well known sources of high energy gamma-rays (see e.g. \cite{par2005}).

In this context, one could think of
this GMRT one-sided jet source as another potential counterpart candidate
for both the HEGRA/Whipple emission deserving further attention in the future. We note here that it could also be consistent
with the HEGRA source when taking into account both the statistical and systematic error quoted in \cite{aha2005b}.
Nevertheless, the extended nature of the HEGRA TeV emission, if confirmed, 
would be difficult to reconcile with a blazar origin for the radiation and therefore such suggestion has
to be considered with caution.

So far, no entry appears catalogued in the SIMBAD database closer than $20^{\prime\prime}$
from the one-sided jet source position in both radio and X-rays. Using the same Chandra data reported by \cite{butt2006},
we find $\sim 10^{-14}$ erg cm$^{-2}$ s$^{-1}$ as a resonable X-ray upper limit at its location within
the 0.3-10 keV energy range.
Such value appears significantly lower than typical X-ray fluxes from other blazar sources, specially when compared with the closest ones
(see e.g. \cite{landt2001}), and thus suggesting a high-redshift object unless beaming is not so strong.

Interestingly, the OMEGA2000 image also displayed in Fig. \ref{gmrt_map_zoom}
reveals a conspicuous stellar-like object
located at $20^h32^m$29\rl 36 and $+41^{\circ}35^{\prime}$06\pri 1, i.e., with an offset
of 1\pri 5 between the radio and near infrared position. This value is a few times the combined
astrometric uncertainty and roughly one third of the GMRT synthesized beam. 
Thus, in principle, a radio/near infrared coincidence should not be considered likely. However, 
an association of the GMRT and CAHA sources cannot be strictly ruled out if the radio peak is not
tracing a core but the photocenter of the much brighter arc-second radio jet next to it and below
the GMRT angular resolution.
From CAHA observations, the magnitude of the near infrared object is $Ks=13.04\pm0.02$. It is also present when
inspecting the 2 Micron All Sky Survey (2MASS) images with similar brightness in the $Ks$ band. 
The 2MASS magnitudes for the candidate counterpart to this GMRT source are
$J=14.77\pm0.03$, $H=13.40\pm0.02$ and $K_s=12.86\pm0.03$.
Higher angular resolution radio observations, as well as near infrared
spectroscopy, is needed to confirm the proposed near infrared counterpart and its
suspected blazar or other nature. 

\subsection{Extended emission}

The multi-configuration VLA 20 cm map in Fig. \ref{vla_c+d_20cm} provides a significantly sharper
image of the diffuse emission inside the HEGRA circle previously reported in \cite{par2007} with half-arc appearance.
The extended emission is broken into multiple patches whose spectral index
is predominantly non-thermal as evidenced in the color coding of Fig. \ref{spix}, thus confirming
the \cite{par2007} estimate. Despite of surrounding the TeV CoG, the morphology of this extended non-thermal
feature does not allow an unambiguous diagnosis of its nature and connection with the gamma-ray
emission, if any. The same problem occurs when inspecting the multi-configuration 6 cm image in Fig. \ref{vla_c+d_6cm}.
At this shorther wavelength, the half-arc becomes more similar to a shell-like feature surrounding the TeV CoG
and the two more compact radio sources close to it.  A similar conclusion has been reached by
\cite{butt2007} when analyzing the D configuration data alone. While these authors tentatively consider
a possible supernova remnant (SNR) reminiscence, this scenario is not clear based on the present data.

\section{Conclusions}

We have conducted a follow-up radio study of the unidentified HEGRA and Whipple TeV sources
in the direction of the Cygnus OB2 association. Our main conclusions can be summarized as follows:

\begin{enumerate}

\item A catalog of relatively compact radio sources in the field has been reported with a total of 153 objects at the 49 cm wavelength
within the primary beam of GMRT antennae. This catalog, available on-line,
is likely to become useful for radio counterpart identication when
the $\gamma$-ray positions are improved in the future and for any other objects in the field.

\item Thanks to the excelent angular resolution of GMRT, at least 6 double radio sources 
are now detected in the field and likely to be double-lobed radio galaxies unrelated to the TeV emission.  
The possibility that the prominent double-lobed source, NVSS J203201+413722,
previously known close to the HEGRA circle and inside the new Whipple ellipse,
is just another unrelated double is therefore enhanced.   

\item The combination of GMRT and VLA maps for NVSS J203201+413722
point toward a double-double lobed radio galaxy nature for this object. A compact optically-thick central component
has been identified with its central core.

\item An interesting one-sided jet radio source has been 
discovered inside the position ellipse of the TeV source detected by Whipple
but could also be consistent with the HEGRA source.
Its radio morphology is reminiscent of an AGN blazar
and it has also a possible stellar-like near infrared counterpart.
In the absence of spectroscopic observations, the physical nature of
this new GMRT source and its connection with the HEGRA/Whipple emission remain possible but certainly unclear. 

\item Finally, the extended radio emission has been also mapped with improved 
detail with the VLA and accurate spectral index maps produced.
A non-thermal shell-like feature around the TeV CoG of TeV~J2032+4130 is confirmed, but a physical association with
the $\gamma$-ray source cannot be established at present in a clear way.
 
\end{enumerate}

\begin{acknowledgements}
{\small The authors  JM, JMP and VBR acknowledge support by
grant AYA2004-07171-C02-01 and AYA2004-07171-C02-02 from the Spanish government, FEDER funds. 
JM is also supported by Plan Andaluz de Investigaci\'on
of Junta de Andaluc\'{\i}a as research group FQM322. 
VBR gratefully acknowledges support from the Alexander von Humboldt 
foundation, and thanks the Max-Planck-Institut f\"ur Kernphysik
for its support and kind hospitality.
GMRT is run by the National Centre for Radio Astrophysics
of the Tata Institute of Fundamental Research.
The NRAO is a facility of the NSF
operated under cooperative agreement by Associated Universities, Inc.
This paper is also based on observations collected at the Centro Astron\'omico Hispano Alem\'an
(CAHA) at Calar Alto, operated jointly by the Max-Planck Institut
f\"ur Astronomie and the Instituto de Astrof\'{\i}sica de
Andaluc\'{\i}a (CSIC). 
This research made use of the SIMBAD
database, operated at the CDS, Strasbourg, France.
This publication makes
use of data products from the Two Micron All Sky Survey, which is a
joint project of the University of Massachusetts and the Infrared
Processing and Analysis Center/California Institute of Technology,
funded by the National Aeronautics and Space Administration and the
National Science Foundation in the USA.
}
\end{acknowledgements}

\Online

\longtab{3}{
\begin{longtable}{ccccccc}
\caption{\label{list} Catalogue of GMRT sources detected at 610 MHz.}\\
\hline\hline
$\alpha_{\rm J2000.0}$ & $\delta_{\rm J2000.0}$  &  $S_{\nu}^{\rm Peak}$ & $S_{\nu}^{\rm Integ}$  &  $a$    &   $b$   &   P.A. \\
   (hms)                &          (dms)          & (mJy beam$^{-1}$) & (mJy)           &  (arcsec)     &   (arcsec)     &   (degree)       \\
\hline
\endfirsthead
\caption{continued.}\\
\hline\hline
 $\alpha_{\rm J2000.0}$ & $\delta_{\rm J2000.0}$  &  $S_{\nu}^{\rm Peak}$ & $S_{\nu}^{\rm Integ}$  &  $a$    &   $b$   &   P.A. \\
   (hms)                &          (dms)          & (mJy beam$^{-1}$) & (mJy)           &  (arcsec)     &   (arcsec)     &   (degree)       \\
\hline
\endhead
\hline
\endfoot
20 30 08.779(0.004) &   +41 33 16.60(0.03) &   8.18(0.11) &  17.30(0.32) &  8.47(0.11) &  5.93(0.08) &  83.0(001) \\
20 30 09.090(0.015) &   +41 33 26.63(0.21) &   2.31(0.10) &   9.23(0.51) & 11.16(0.51) &  8.51(0.39) &   4.0(007) \\
20 30 11.448(0.015) &   +41 34 52.19(0.15) &   2.14(0.11) &   4.95(0.34) &  8.29(0.42) &  6.63(0.33) &  68.5(009) \\
20 30 13.065(0.066) &   +41 35 29.88(0.59) &   0.52(0.11) &   0.94(0.29) &  9.58(2.02) &  4.51(0.95) &  54.5(010) \\
20 30 14.295(0.040) &   +41 24 54.89(0.52) &   0.66(0.11) &   1.30(0.30) &  7.48(1.24) &  6.26(1.03) &  18.5(037) \\
20 30 17.861(0.068) &   +41 30 12.04(0.64) &   0.46(0.11) &   0.74(0.27) &  8.85(2.11) &  4.31(1.03) & 126.9(012) \\
20 30 19.401(0.040) &   +41 24 23.69(0.58) &   0.60(0.11) &   1.10(0.29) &  7.47(1.36) &  5.83(1.06) &   1.0(029) \\
20 30 22.154(0.015) &   +41 28 10.15(0.14) &   1.85(0.11) &   2.63(0.25) &  6.90(0.42) &  4.89(0.29) &  65.8(007) \\
20 30 22.502(0.054) &   +41 39 21.53(0.38) &   0.64(0.11) &   1.12(0.28) &  8.32(1.44) &  5.03(0.87) &  98.2(013) \\
20 30 23.054(0.034) &   +41 39 48.70(0.63) &   0.50(0.11) &   0.54(0.21) &  6.52(1.49) &  3.90(0.89) & 178.4(017) \\
20 30 24.657(0.010) &   +41 24 45.81(0.10) &   2.61(0.11) &   3.81(0.25) &  7.08(0.30) &  4.89(0.21) &  57.4(005) \\
20 30 28.735(0.070) &   +41 19 15.41(0.47) &   0.57(0.11) &   1.37(0.35) & 10.11(1.89) &  5.61(1.05) & 103.9(012) \\
20 30 29.748(0.032) &   +41 18 20.76(0.60) &   0.60(0.11) &   0.83(0.24) &  7.72(1.44) &  4.25(0.80) & 166.7(012) \\
20 30 34.471(0.036) &   +41 44 18.58(0.53) &   0.63(0.11) &   1.01(0.26) &  7.14(1.25) &  5.32(0.93) &   8.2(024) \\
20 30 37.252(0.036) &   +41 18 28.27(0.73) &   0.47(0.11) &   0.54(0.22) &  7.09(1.72) &  3.88(0.94) & 178.6(015) \\
20 30 37.708(0.046) &   +41 18 15.10(0.58) &   0.49(0.11) &   0.55(0.21) &  6.61(1.55) &  4.06(0.95) &  37.3(019) \\
20 30 37.787(0.031) &   +41 31 44.99(0.58) &   0.52(0.11) &   0.46(0.18) &  6.32(1.40) &  3.35(0.74) & 162.3(013) \\
20 30 38.443(0.030) &   +41 33 33.53(0.42) &   0.78(0.11) &   1.28(0.27) &  6.99(0.98) &  5.57(0.79) & 171.1(025) \\
20 30 39.671(0.044) &   +41 32 19.40(0.70) &   0.49(0.11) &   0.76(0.26) &  7.39(1.68) &  5.00(1.13) & 169.1(023) \\
20 30 40.094(0.001) &   +41 23 30.71(0.01) & 228.61(0.11) & 326.27(0.25) &  7.03(0.00) &  4.82(0.00) &  57.5(000) \\
20 30 42.997(0.050) &   +41 41 11.50(0.69) &   0.51(0.11) &   0.68(0.24) &  8.70(1.93) &  3.69(0.82) & 143.9(009) \\
20 30 44.365(0.057) &   +41 26 47.74(0.64) &   0.54(0.11) &   1.26(0.34) &  8.24(1.65) &  6.75(1.35) &  46.1(040) \\
20 30 45.785(0.045) &   +41 41 11.04(0.83) &   0.49(0.11) &   0.94(0.30) &  8.73(1.95) &  5.27(1.18) & 178.7(017) \\
20 30 46.058(0.039) &   +41 19 37.85(0.54) &   0.53(0.11) &   0.63(0.22) &  6.11(1.31) &  4.61(0.99) & 157.8(030) \\
20 30 48.436(0.047) &   +41 19 52.89(0.60) &   0.47(0.11) &   0.57(0.22) &  6.41(1.54) &  4.51(1.08) &  35.2(027) \\
20 30 49.742(0.026) &   +41 36 12.76(0.28) &   1.05(0.11) &   1.79(0.28) &  6.67(0.70) &  6.07(0.64) & 115.1(045) \\
20 30 52.837(0.033) &   +41 15 08.57(0.34) &   0.60(0.11) &   0.48(0.17) &  4.71(0.90) &  4.07(0.78) &  62.7(053) \\
20 30 55.876(0.046) &   +41 19 04.57(0.53) &   0.47(0.11) &   0.48(0.20) &  5.71(1.39) &  4.29(1.04) &  43.1(034) \\
20 30 56.254(0.040) &   +41 25 32.54(0.71) &   0.63(0.11) &   1.33(0.32) & 10.34(1.79) &  4.90(0.85) &  23.2(009) \\
20 31 01.197(0.040) &   +41 47 46.30(0.63) &   0.51(0.11) &   0.50(0.20) &  7.33(1.66) &  3.19(0.72) &  30.4(010) \\
20 31 01.918(0.026) &   +41 14 23.73(0.67) &   0.54(0.11) &   0.55(0.20) &  7.42(1.57) &  3.27(0.69) &   1.4(009) \\
20 31 03.227(0.038) &   +41 45 40.33(0.55) &   0.52(0.11) &   0.46(0.18) &  6.71(1.48) &  3.16(0.70) & 145.9(011) \\
20 31 03.846(0.009) &   +41 26 42.57(0.10) &   2.66(0.11) &   3.50(0.23) &  6.31(0.27) &  4.95(0.21) &  55.2(007) \\
20 31 04.137(0.031) &   +41 36 50.59(0.43) &   0.62(0.11) &   0.63(0.20) &  5.45(1.01) &  4.43(0.82) & 174.9(036) \\
20 31 07.944(0.032) &   +41 14 05.72(0.29) &   0.95(0.11) &   1.39(0.25) &  8.09(0.95) &  4.31(0.51) &  56.2(007) \\
20 31 08.779(0.033) &   +41 14 07.10(0.34) &   0.86(0.11) &   1.18(0.24) &  8.13(1.06) &  4.02(0.53) &  49.6(007) \\
20 31 10.552(0.040) &   +41 30 20.46(0.51) &   0.56(0.11) &   0.60(0.21) &  7.01(1.43) &  3.61(0.74) &  39.0(012) \\
20 31 10.641(0.052) &   +41 29 59.63(0.65) &   0.46(0.11) &   0.72(0.26) &  6.52(1.55) &  5.67(1.35) &  14.2(069) \\
20 31 12.212(0.002) &   +41 39 04.60(0.02) &  12.06(0.11) &  19.34(0.27) &  6.69(0.06) &  5.69(0.05) & 130.4(002) \\
20 31 13.212(0.002) &   +41 39 01.87(0.02) &  15.51(0.11) &  24.80(0.27) &  6.55(0.05) &  5.79(0.04) & 105.3(002) \\
20 31 13.360(0.043) &   +41 17 06.93(0.68) &   0.53(0.11) &   0.77(0.25) &  8.46(1.77) &  4.08(0.85) &  29.2(011) \\
20 31 14.174(0.028) &   +41 38 42.69(0.34) &   0.81(0.11) &   1.01(0.23) &  6.22(0.87) &  4.77(0.67) &  35.8(021) \\
20 31 15.201(0.007) &   +41 11 23.90(0.10) &   3.56(0.11) &   5.83(0.27) &  8.07(0.25) &  4.83(0.15) &  30.4(002) \\
20 31 15.643(0.023) &   +41 34 19.51(0.28) &   0.86(0.11) &   0.83(0.19) &  5.05(0.67) &  4.53(0.60) &  24.8(050) \\
20 31 16.882(0.016) &   +41 49 31.97(0.35) &   0.78(0.11) &   0.52(0.15) &  5.61(0.82) &  2.82(0.41) &   1.7(008) \\
20 31 19.038(0.054) &   +41 34 42.16(0.55) &   0.56(0.11) &   1.17(0.32) &  7.44(1.44) &  6.65(1.29) &  90.8(070) \\
20 31 21.132(0.021) &   +41 34 59.74(0.27) &   0.97(0.11) &   1.04(0.21) &  5.43(0.64) &  4.67(0.55) & 175.0(031) \\
20 31 24.954(0.052) &   +41 50 24.53(0.78) &   0.48(0.11) &   0.98(0.31) &  8.40(1.89) &  5.75(1.29) & 159.9(023) \\
20 31 25.132(0.037) &   +41 49 26.34(0.64) &   0.51(0.11) &   0.50(0.20) &  7.23(1.63) &  3.22(0.73) & 154.2(010) \\
20 31 25.968(0.072) &   +41 46 18.73(0.80) &   0.47(0.11) &   1.13(0.35) & 10.16(2.33) &  5.63(1.29) &  45.1(015) \\
20 31 29.616(0.013) &   +41 23 36.68(0.15) &   1.77(0.11) &   2.19(0.23) &  6.12(0.39) &  4.80(0.31) &  42.8(011) \\
20 31 29.784(0.036) &   +41 12 07.81(0.55) &   0.66(0.11) &   0.96(0.25) &  8.85(1.49) &  3.89(0.65) &  32.2(007) \\
20 31 32.243(0.065) &   +41 47 34.13(1.11) &   0.51(0.11) &   1.55(0.41) & 14.14(2.94) &  5.10(1.06) & 150.7(007) \\
20 31 32.448(0.019) &   +41 47 56.61(0.25) &   1.40(0.11) &   2.83(0.31) &  7.58(0.59) &  6.35(0.49) & 174.1(018) \\
20 31 36.759(0.025) &   +41 17 40.36(0.37) &   0.80(0.11) &   0.85(0.21) &  6.55(0.94) &  3.87(0.56) &  29.2(011) \\
20 31 39.166(0.018) &   +41 18 19.15(0.26) &   0.95(0.11) &   0.81(0.18) &  5.19(0.62) &  3.91(0.47) & 169.4(017) \\
20 31 44.650(0.017) &   +41 51 34.76(0.23) &   1.20(0.11) &   1.37(0.22) &  5.80(0.55) &  4.68(0.45) & 176.3(018) \\
20 31 47.002(0.003) &   +41 52 12.80(0.05) &   8.01(0.11) &  14.40(0.29) &  7.91(0.11) &  5.40(0.07) & 172.5(001) \\
20 31 51.782(0.020) &   +41 31 18.29(0.27) &   1.15(0.11)$^a$ &   1.73(0.26) &  6.47(0.62) &  5.52(0.53) &   1.3(025) \\
20 31 58.528(0.044) &   +41 52 25.72(0.73) &   0.54(0.11) &   1.10(0.31) &  8.53(1.72) &  5.68(1.15) & 176.5(020) \\
20 32 00.121(0.044) &   +41 16 47.40(0.64) &   0.51(0.11) &   0.71(0.24) &  7.31(1.61) &  4.54(1.00) &  28.2(018) \\
20 32 01.094(0.041) &   +41 49 59.15(0.48) &   0.54(0.11) &   0.59(0.21) &  5.90(1.25) &  4.39(0.93) & 140.6(028) \\
20 32 03.399(0.053) &   +41 37 38.40(0.67) &   0.68(0.10) &   2.58(0.49) & 10.67(1.65) &  8.45(1.31) & 148.8(027) \\
20 32 04.062(0.046) &   +41 46 37.39(0.64) &   0.49(0.11) &   0.69(0.25) &  6.75(1.55) &  5.00(1.15) &  21.0(031) \\
20 32 04.938(0.023) &   +41 42 35.83(0.30) &   0.94(0.11) &   1.15(0.22) &  6.11(0.73) &  4.74(0.57) & 156.5(019) \\
20 32 06.183(0.023) &   +41 14 05.73(0.31) &   0.96(0.11) &   1.31(0.24) &  6.73(0.78) &  4.80(0.56) & 150.4(014) \\
20 32 07.178(0.008) &   +41 48 38.47(0.11) &   2.72(0.11) &   3.46(0.23) &  6.23(0.26) &  4.85(0.20) &  15.1(007) \\
20 32 07.337(0.010) &   +41 37 26.73(0.09) &   2.93(0.11) &   4.82(0.27) &  7.01(0.26) &  5.56(0.21) &  88.0(007) \\
20 32 08.365(0.032) &   +41 13 11.72(0.52) &   0.65(0.11) &   0.96(0.25) &  7.40(1.26) &  4.72(0.80) &  14.9(015) \\
20 32 08.401(0.013) &   +41 43 22.65(0.19) &   1.66(0.11) &   2.33(0.24) &  6.85(0.46) &  4.85(0.33) &  12.2(008) \\
20 32 09.435(0.014) &   +41 35 57.89(0.17) &   1.57(0.11) &   1.94(0.23) &  5.57(0.40) &  5.26(0.38) & 174.5(051) \\
20 32 09.550(0.039) &   +41 38 06.36(0.82) &   0.51(0.11) &   0.93(0.29) &  8.95(1.93) &  4.84(1.04) & 178.8(013) \\
20 32 10.985(0.038) &   +41 32 39.57(0.48) &   0.52(0.11)$^b$ &   0.51(0.20) &  5.36(1.18) &  4.32(0.95) & 154.5(041) \\
20 32 12.738(0.016) &   +41 49 28.16(0.28) &   1.27(0.11) &   1.89(0.25) &  7.43(0.65) &  4.75(0.42) & 172.6(008) \\
20 32 13.092(0.039) &   +41 27 24.16(0.48) &   0.51(0.11) &   0.46(0.19) &  5.88(1.31) &  3.65(0.81) & 140.7(018) \\
20 32 13.292(0.070) &   +41 13 51.21(0.90) &   0.60(0.10) &   3.24(0.66) & 12.51(2.17) & 10.32(1.79) &  24.1(036) \\
20 32 13.681(0.040) &   +41 15 03.13(0.67) &   0.48(0.11) &   0.52(0.21) &  7.18(1.71) &  3.55(0.85) & 153.9(013) \\
20 32 14.291(0.012) &   +41 16 36.03(0.19) &   1.90(0.11) &   3.11(0.27) &  8.17(0.47) &  4.75(0.28) & 152.2(004) \\
20 32 15.361(0.032) &   +41 22 19.01(0.46) &   0.64(0.11) &   0.78(0.22) &  6.30(1.11) &  4.54(0.80) & 160.2(021) \\
20 32 15.595(0.017) &   +41 13 28.27(0.26) &   1.35(0.11) &   2.17(0.27) &  7.89(0.65) &  4.84(0.40) & 154.6(007) \\
20 32 16.289(0.043) &   +41 30 55.57(0.48) &   0.57(0.11)$^c$ &   0.77(0.24) &  6.04(1.20) &  5.37(1.06) &  44.2(069) \\
20 32 16.509(0.051) &   +41 49 45.75(0.64) &   0.49(0.11) &   0.62(0.23) &  7.81(1.81) &  3.88(0.90) & 140.5(012) \\
20 32 18.936(0.048) &   +41 08 58.18(1.00) &   0.47(0.11) &   1.11(0.35) & 10.24(2.35) &  5.52(1.26) & 177.1(014) \\
20 32 19.879(0.050) &   +41 11 57.99(0.71) &   0.48(0.11) &   0.83(0.28) &  7.87(1.79) &  5.20(1.18) &  27.4(022) \\
20 32 21.023(0.054) &   +41 34 21.21(0.49) &   0.52(0.11) &   0.77(0.25) &  6.63(1.42) &  5.31(1.14) &  80.7(039) \\
20 32 21.296(0.040) &   +41 09 14.86(0.69) &   0.61(0.11) &   1.24(0.31) &  9.77(1.73) &  4.89(0.87) &  23.7(010) \\
20 32 22.487(0.011) &   +41 18 17.80(0.13) &   2.07(0.11)$^d$ &   2.41(0.22) &  5.51(0.30) &  5.03(0.28) & 150.9(024) \\
20 32 23.349(0.059) &   +41 49 14.67(0.68) &   0.47(0.11) &   0.70(0.25) &  8.60(2.01) &  4.09(0.96) & 136.6(012) \\
20 32 27.113(0.043) &   +41 41 28.64(0.67) &   0.48(0.11) &   0.64(0.24) &  7.11(1.65) &  4.42(1.03) &  21.3(019) \\
20 32 28.252(0.051) &   +41 14 55.08(0.47) &   0.50(0.11) &   0.56(0.21) &  6.43(1.47) &  4.12(0.94) &  60.1(020) \\
20 32 28.647(0.007) &   +41 42 07.61(0.08) &   3.45(0.11) &   4.57(0.24) &  6.21(0.20) &  5.08(0.17) &  27.7(007) \\
20 32 28.943(0.060) &   +41 16 40.68(0.84) &   0.46(0.11) &   0.65(0.24) &  9.91(2.40) &  3.34(0.81) &  37.3(007) \\
20 32 29.477(0.009) &   +41 35 06.67(0.09) &   2.82(0.11) &   3.73(0.24) &  5.72(0.23) &  5.49(0.22) & 121.7(038) \\
20 32 32.739(0.029) &   +41 46 30.51(0.38) &   0.75(0.11) &   0.94(0.23) &  6.13(0.93) &  4.89(0.74) &  23.1(027) \\
20 32 32.982(0.040) &   +41 35 01.41(0.45) &   0.59(0.11) &   0.79(0.24) &  5.70(1.08) &  5.54(1.05) &  28.2(267) \\
20 32 34.141(0.007) &   +41 40 42.63(0.09) &   3.13(0.11) &   4.00(0.23) &  5.93(0.21) &  5.11(0.18) &  21.2(010) \\
20 32 34.286(0.040) &   +41 24 12.57(0.39) &   0.75(0.11) &   1.45(0.30) &  7.31(1.06) &  6.28(0.91) &  98.4(038) \\
20 32 36.683(0.007) &   +41 14 45.84(0.10) &   3.46(0.11) &   5.55(0.27) &  7.08(0.23) &  5.38(0.17) & 170.9(005) \\
20 32 37.211(0.056) &   +41 14 53.48(0.77) &   0.67(0.10) &   2.56(0.49) & 13.39(2.10) &  6.76(1.06) & 145.3(009) \\
20 32 38.196(0.001) &   +41 23 36.67(0.01) &  19.32(0.11) &  26.15(0.24) &  6.15(0.04) &  5.23(0.03) & 167.9(001) \\
20 32 38.754(0.030) &   +41 23 13.81(0.82) &   0.47(0.11) &   0.52(0.21) &  8.00(1.93) &  3.23(0.78) &   2.5(009) \\
20 32 40.051(0.089) &   +41 21 25.22(0.88) &   0.46(0.11) &   1.14(0.36) & 12.47(2.93) &  4.78(1.12) & 129.8(009) \\
20 32 40.997(0.025) &   +41 14 27.76(0.34) &   0.82(0.11)$^e$ &   0.93(0.22) &  5.73(0.80) &  4.71(0.66) & 168.1(029) \\
20 32 43.982(0.036) &   +41 19 08.80(0.58) &   0.55(0.11) &   0.67(0.22) &  7.02(1.44) &  4.12(0.85) & 158.5(015) \\
20 32 44.630(0.008) &   +41 32 46.85(0.09) &   2.91(0.11) &   3.51(0.22) &  5.61(0.22) &  5.11(0.20) &  46.7(017) \\
20 32 44.753(0.041) &   +41 12 20.82(0.66) &   0.50(0.11) &   0.53(0.21) &  7.52(1.73) &  3.34(0.77) &  30.7(010) \\
20 32 45.367(0.013) &   +41 39 22.37(0.16) &   1.64(0.11) &   1.78(0.21) &  5.29(0.37) &  4.88(0.34) &  12.8(035) \\
20 32 49.093(0.031) &   +41 10 52.54(0.34) &   0.66(0.11) &   0.56(0.18) &  5.58(0.96) &  3.59(0.62) &  46.4(015) \\
20 32 51.789(0.051) &   +41 10 56.85(0.62) &   0.51(0.11) &   0.66(0.23) &  8.12(1.81) &  3.81(0.85) &  42.4(011) \\
20 32 51.980(0.083) &   +41 46 56.44(1.02) &   0.46(0.10) &   1.95(0.54) & 11.04(2.50) &  9.11(2.06) & 149.2(048) \\
20 32 52.081(0.038) &   +41 45 00.20(0.58) &   0.47(0.11) &   0.46(0.20) &  5.60(1.37) &  4.13(1.01) & 175.1(032) \\
20 32 52.697(0.048) &   +41 17 21.25(0.69) &   0.50(0.11) &   0.71(0.25) &  8.42(1.87) &  4.00(0.89) &  33.7(011) \\
20 32 52.968(0.051) &   +41 48 11.71(0.71) &   0.46(0.11) &   0.76(0.27) &  6.97(1.67) &  5.61(1.35) & 178.7(045) \\
20 32 53.082(0.038) &   +41 34 31.10(0.53) &   0.88(0.11) &   1.98(0.33) & 12.45(1.52) &  4.27(0.52) &  37.2(004) \\
20 32 53.129(0.060) &   +41 10 49.45(0.56) &   0.50(0.11) &   0.72(0.25) &  8.23(1.84) &  4.20(0.94) &  54.3(012) \\
20 32 54.292(0.052) &   +41 35 02.24(0.58) &   0.62(0.11) &   1.48(0.35) &  9.48(1.63) &  5.93(1.02) &  45.9(014) \\
20 32 58.046(0.088) &   +41 20 43.94(1.00) &   0.48(0.10) &   2.23(0.58) & 11.92(2.60) &  9.30(2.03) &  44.2(035) \\
20 33 00.408(0.095) &   +41 40 59.42(1.13) &   0.46(0.10) &   2.60(0.68) & 11.93(2.67) & 11.15(2.49) &  10.0(133) \\
20 33 04.046(0.070) &   +41 14 05.15(1.07) &   0.48(0.11) &   1.03(0.32) & 13.21(3.00) &  3.89(0.89) &  34.6(006) \\
20 33 05.432(0.041) &   +41 26 00.92(0.41) &   0.69(0.11) &   1.19(0.28) &  6.83(1.09) &  5.99(0.95) &  97.0(049) \\
20 33 07.879(0.079) &   +41 18 06.28(0.67) &   0.47(0.11) &   1.23(0.37) &  9.44(2.15) &  6.60(1.50) & 107.6(025) \\
20 33 10.760(0.015) &   +41 15 05.93(0.21) &   1.25(0.11)$^f$ &   1.24(0.20) &  5.42(0.50) &  4.35(0.40) & 167.6(017) \\
20 33 11.272(0.001) &   +41 42 09.41(0.01) &  24.17(0.11) &  42.79(0.28) &  7.17(0.03) &  5.86(0.03) &  55.2(001) \\
20 33 11.771(0.053) &   +41 48 22.76(0.91) &   0.49(0.11) &   1.19(0.35) & 10.26(2.24) &  5.62(1.23) & 160.1(014) \\
20 33 12.374(0.034) &   +41 47 30.61(0.41) &   0.63(0.11) &   0.69(0.21) &  5.46(0.99) &  4.76(0.87) &  23.3(054) \\
20 33 12.622(0.051) &   +41 15 34.80(0.75) &   0.49(0.11) &   0.80(0.27) &  8.85(2.00) &  4.38(0.99) &  32.8(012) \\
20 33 13.248(0.076) &   +41 15 29.37(0.73) &   0.57(0.10) &   2.34(0.53) & 11.82(2.18) &  8.26(1.52) & 121.5(020) \\
20 33 15.139(0.012) &   +41 28 21.08(0.09) &   2.41(0.11) &   3.06(0.23) &  6.62(0.31) &  4.56(0.21) &  97.7(005) \\
20 33 19.499(0.033) &   +41 15 22.98(0.63) &   0.53(0.11) &   0.62(0.22) &  6.90(1.48) &  4.05(0.87) &   2.3(016) \\
20 33 19.632(0.004) &   +41 28 30.71(0.04) &   6.07(0.11) &   7.91(0.23) &  6.01(0.11) &  5.15(0.10) & 118.0(005) \\
20 33 19.806(0.055) &   +41 32 33.97(0.30) &   0.92(0.11) &   3.08(0.44) & 12.72(1.45) &  6.22(0.71) &  92.9(006) \\
20 33 21.157(0.080) &   +41 32 31.85(0.68) &   0.46(0.11) &   1.16(0.36) &  8.97(2.11) &  6.75(1.58) &  86.1(033) \\
20 33 22.784(0.045) &   +41 18 45.37(0.65) &   0.54(0.11) &   0.94(0.28) &  7.94(1.62) &  5.23(1.07) &  25.8(019) \\
20 33 24.301(0.014) &   +41 34 06.05(0.15) &   1.59(0.11) &   1.65(0.20) &  5.21(0.38) &  4.75(0.34) &  78.8(031) \\
20 33 28.186(0.034) &   +41 45 58.07(0.43) &   0.65(0.11) &   0.76(0.22) &  6.45(1.13) &  4.29(0.75) &  37.3(017) \\
20 33 30.962(0.024) &   +41 35 27.45(0.23) &   0.94(0.11) &   0.89(0.19) &  5.73(0.70) &  3.92(0.48) &  58.6(013) \\
20 33 31.553(0.047) &   +41 29 59.72(0.69) &   0.49(0.11) &   0.62(0.23) &  7.84(1.82) &  3.85(0.89) &  32.6(012) \\
20 33 33.307(0.046) &   +41 16 20.50(0.57) &   0.68(0.11) &   1.94(0.39) &  8.75(1.36) &  7.72(1.20) &  19.1(050) \\
20 33 33.370(0.055) &   +41 35 52.10(0.58) &   0.47(0.11) &   0.65(0.24) &  6.61(1.59) &  5.00(1.20) &  51.7(034) \\
20 33 36.133(0.023) &   +41 42 54.02(0.31) &   0.81(0.11) &   0.60(0.16) &  5.91(0.83) &  2.99(0.42) &  35.8(008) \\
20 33 38.345(0.041) &   +41 41 08.39(0.64) &   0.48(0.11) &   0.57(0.22) &  6.61(1.55) &  4.26(1.00) &  20.5(021) \\
20 33 38.376(0.010) &   +41 39 54.51(0.09) &   3.00(0.11) &   4.58(0.26) &  7.58(0.28) &  4.77(0.18) &  60.1(003) \\
20 33 39.878(0.010) &   +41 38 30.08(0.10) &   2.57(0.11) &   3.71(0.25) &  6.31(0.27) &  5.42(0.23) &  87.5(011) \\
20 33 40.681(0.033) &   +41 40 25.78(0.60) &   0.58(0.11) &   0.59(0.20) &  7.80(1.55) &  3.13(0.62) &  25.6(008) \\
20 33 40.761(0.069) &   +41 27 02.77(0.81) &   0.64(0.10) &   3.59(0.68) & 12.02(1.94) & 11.07(1.79) &  34.7(079) \\
20 33 41.453(0.049) &   +41 26 30.60(0.61) &   0.61(0.11) &   1.51(0.36) &  8.54(1.50) &  6.93(1.22) &  30.3(034) \\
20 33 44.630(0.037) &   +41 40 22.77(0.52) &   1.15(0.10) &   7.02(0.73) & 14.01(1.26) & 10.34(0.93) &  22.3(012) \\
20 33 50.746(0.032) &   +41 29 16.47(0.52) &   0.65(0.11) &   0.70(0.21) &  7.59(1.34) &  3.40(0.60) &  29.0(008) \\
20 33 52.064(0.020) &   +41 21 50.48(0.23) &   1.60(0.11) &   4.36(0.38) &  8.30(0.55) &  7.77(0.52) & 146.1(040) \\
20 33 59.042(0.055) &   +41 27 41.55(0.66) &   0.51(0.11) &   1.09(0.32) &  7.50(1.58) &  6.72(1.42) & 151.1(077) \\
20 33 59.598(0.063) &   +41 25 39.82(0.89) &   0.53(0.11) &   1.90(0.47) & 11.18(2.21) &  7.57(1.49) &  26.9(020) \\
20 33 59.649(0.013) &   +41 35 36.36(0.11) &   2.46(0.11) &   4.61(0.29) &  7.69(0.34) &  5.79(0.26) &  91.8(006) \\
20 34 02.483(0.014) &   +41 24 54.50(0.12) &   2.28(0.11) &   4.21(0.29) &  7.67(0.37) &  5.71(0.27) & 105.2(006) \\
\end{longtable}
\noindent
$^a$ Same as source \#3, VLA-North and Chandra \#234 within the HEGRA circle in \cite{par2007}.\\
$^b$ Same as source \#4 within the HEGRA circle in \cite{par2007}.\\
$^c$ Same as source \#6 within the HEGRA circle in \cite{par2007}.\\
$^d$ Likely the 610 MHz radio counterpart of the contact binary system Cyg OB2 No. 5.\\
$^e$ Likely the 610 MHz radio counterpart of the luminous star Cyg OB2 No. 12.\\
$^f$ Likely the 610 MHz radio counterpart of the luminous star Cyg OB2 No. 9.\\
}

\end{document}